\documentstyle[preprint,aps,prb]{revtex}
\begin{document}
\draft
\preprint{to appear in \  J.\ Phys.\ Soc.\ Jpn.\ {\bf 66}, No.\ 5 (1997)} 
\title{ 
       Fermi-liquid theory for a conductance through an  
       interacting region attached to noninteracting leads
}
\author{Akira Oguri}
\address{
          Department of Material Science,
          Faculty of Science,
          Osaka City University, \\
          Sumiyoshi-ku, Osaka 558,
          Japan
}

\date{Received \ January 8, 1997}
\maketitle

\begin{abstract}
We study the conductance for the current through 
a finite interacting, possibly disordered region 
attached to two noninteracting leads within the Kubo formalism
applying the perturbation theory for the local Fermi liquid  
developed by Yamada \& Yosida and by Shiba to this finite region. 
Assuming the validity of the perturbation expansion 
in the Coulomb interaction and 
the time reversal symmetry for the normal scattering potential 
in the finite region, 
we find that the contributions of the vertex correction 
to the dc conductance disappear at $\,T=0$ 
if the currents are measured in the leads.
Consequently, 
the conductance is expressed as 
$g=(2 e^2/h)\, |t(0)|^2$. Here $t(0)$ is
the transmission coefficient  
for single-particle-like excitation at the Fermi energy  
and is introduced by using the Green's function. 
The results are generalized to a quasi-one-dimensional system 
with a number of scattering channels.
\end{abstract}
\bigskip
\bigskip


\narrowtext

\section{INTRODUCTION}
\label{sec:intro}
The effect of the Coulomb interaction 
on the process of resonant tunneling through 
a small region has been a subject of considerable investigation 
in the past few years. 
For studying the transport through such small systems theoretically, 
it is sometimes necessary  
to formulate the expression of the conductance 
so that both the Coulomb interaction 
and quantum mechanical interference effects 
can be treated exactly, and  
several attempts to obtain a rather general formulation 
have been made so far.\cite{Bee,MW}
The purpose of this paper is 
to present one such attempt based on the Kubo formalism 
with the perturbation theory in the Coulomb interaction.

We consider a system in which the Coulomb interaction 
and the scattering potential due to the disorder
are switched on only for the electrons staying 
in a finite region of an infinite system, 
and apply the perturbation analysis 
used in the microscopic version of the Fermi liquid theory 
for the impurity Anderson model\cite{YY3,Yamada4,Shiba,Yoshimori}
to the present system. 
We find that the dc conductance for zero-temperature $\,T=0\,$ 
is written in a Landauer-type form\cite{Landauer,Buttiker} 
if the assumptions shown in the following are satisfied.
This result may be regarded as an extension of 
the relation shown in the noninteracting case  
by Economou \& Soukoulis for one dimension and 
by Fisher \& Lee for higher dimensions.\cite{Economou,Fisher,Lee} 
The assumptions to be made are 
\begin{description}
\item[$\phantom{ii}i$)] {\em 
the Coulomb interaction and 
the normal spin-independent scattering potential are 
restricted for electrons in a finite spatial region, and
two semi-infinite noninteracting leads are coupled to
this region\/},
\item[$\phantom{i}ii$)] 
{\em perturbation expansion in the Coulomb interaction   
is valid\/},
\item[$iii$)] \ {\em the single-particle-like excitation 
at the chemical potential $\mu$ does not decay,
i.e., the imaginary part of the proper self-energy 
which is no longer diagonal in the wave-number indices 
is zero at $\,T=0$\/}.
\end{description}
The assumption $i$) specifies the model, and 
means that the Hamiltonian does not contain a 
spin-flip-scattering due to the Kondo-type exchange interaction, 
the spin-orbit coupling, and so on.
The noninteracting leads make 
the scattering problem of the present system well defined, 
and enable us to introduce the scattering coefficients 
making use of the asymptotic form of the single-particle Green's function 
in the leads. 
Furthermore, the presence of the semi-infinite leads 
seems to make the Fermi-liquid description
for low-energy states, the wave-length of which is much longer  
than the size of the central region $\,L\,$, probable.
Under the assumption $ii$),  
we use the diagrammatic analysis 
for the perturbation in the Coulomb interaction 
considering the central region as {\em an impurity\/} 
with a finite volume,\cite{YY3,Yamada4,Shiba}
and calculate the dc conductance by extracting 
the $\omega$-linear imaginary part of 
the current-current correlation function following Shiba.\cite{Shiba} 
The assumption $\,iii$) was originally introduced 
by Langer \& Ambegaokar as a basic assumption in deriving 
the Friedel sum rule for interacting Fermi systems.
\cite{LangerAmbegaokar}
As Langer \& Ambegaokar also mentioned,
this assumption holds in the perturbation theory.   
It means that $\,iii$) is not an independent assumption
and deduced from the two assumptions $\,i$) and $ii$). 
We will show this  explicitly by calculating 
the imaginary part of the self-energy following Yamada \& Yosida.
\cite{YY3,Yamada4}

In this paper we consider mainly the system in one dimension. 
This is because the extension to higher dimensions 
is possible following along almost the same lines. 
We show the outline of the extension in the last part of this paper.
Owing to the presence of the noninteracting leads,
the perturbation theory will probably make sense 
for low-energy states with long enough wave-length 
even in the one-dimensional case.
The Tomonaga-Luttinger-liquid-like behavior 
seems to be valid for {\em higher-energy\/} region where 
the wave-length of a massless-boson-like excitation is shorter 
than the size of the central region $L$.\cite{KaneFisher,FurusakiNagaosa} 
Recently, 
the value of the renormalization factor $K$ of the conductance  
in the Tomonaga-Luttinger liquid, 
$g= K e^2/h$ (per spin), was discussed by several authors
\cite{MaslovStone,Ponomarenko,Kawabata2,Shimizu}
in relation to the experiment by Tarucha, Honda, and Saku.\cite{Tarucha}
In the perturbation theory, 
the conductance can be expressed in a Landauer-type form, 
so that $\,g$ becomes $e^2/h$ (per spin) 
when a resonant tunneling occurs, i.e., 
in the case the transmission coefficient for the single-particle-like 
excitation at the Fermi energy is unity. 
This result itself seems to be consistent with 
the result obtained by Shimizu with the phenomenological description of 
a Fermi liquid in one dimension.\cite{Shimizu}

We note that some of the ideas used in the present study were  
already applied to the current through 
a single Anderson-Wolff impurity\cite{AndersonModel,Wolff} 
in our previous work,\cite{ao3}
where the expression of the conductance was studied
for the use of numerical calculations 
with the quantum Monte Carlo method.\cite{ao,ao4} 
The expression for the conductance through the single impurity
has also been studied precisely with the Keldysh formalism
by Hersfield, Davies and Willkins.\cite{HDW2}
The conductance through 
a two-impurity Anderson-Wolff model 
has been formulated by Izumida, Sakai and Shimizu 
with a linear response theory 
for the use of the calculations with the numerical 
renormalization group method.\cite{Izumida}

In section \ref{sec:trans}, we introduce the 
transmission and reflection coefficients, 
and confirm that the coefficients 
for the single-particle-like excitation at the Fermi energy 
satisfy the unitarity condition.
In section \ref{sec:self-energy}, 
we calculate the imaginary part of the proper self-energy 
with the perturbation theory in the Coulomb interaction,
and show the assumption $\,iii$) is deduced from $\,i$) and $ii$).
In section \ref{sec:Kubo}, 
we study the expression for the dc conductance $g$ within the Kubo formalism, 
and show that the contributions of the vertex correction to $g$ are zero 
at $T=0$ if the currents are measured in the noninteracting leads.
Owing to this property, $g$ can be expressed 
in terms of the transmission coefficient. 
We also discuss the role the vertex correction 
with relation to the back flow.
In section \ref{sec:extension}, 
we generalize the results obtained for one dimension
in Sec.\  \ref{sec:trans}--\ref{sec:Kubo} 
to the quasi-one-dimensional system 
which is finite along the transverse direction.

\section{SCATTERING COEFFICIENTS}
\label{sec:trans}

In this section, we introduce the 
transmission and reflection coefficients 
for a single-particle-like excitation, 
and show that the coefficients 
for the excitation at the Fermi energy 
satisfy the unitarity condition 
if the assumption $\,iii)$ holds.

We consider an one-dimensional system
which consists of three parts: 
a finite central region at $0 <  x < L$, 
and two semi-infinite lead wires 
at $\,-\infty < x \leq 0\,$ and $\,L \leq x < +\infty\,$. 
The Hamiltonian is given by
\begin{eqnarray}
 {\cal H}_{\phantom{0}} &=& \ {\cal H}_0 + {\cal H}_V + {\cal H}_U 
\;, 
\label{eq:H}
\\
 {\cal H}_0 &=& \  
  \sum_{\sigma} \int_{-\infty}^{+\infty}\!\!  dx\ 
  \psi^{\dagger}_{\sigma}(x)\,
   \left( - {1 \over 2m} {\partial^2 \over \partial x^2} -\mu \right) 
   \,\psi^{\phantom{\dagger}}_{\sigma}(x) \;,
   \nonumber \\
  {\cal H}_V &=& \ 
\sum_{\sigma} \int_{0}^{L}\!\!  dx\ 
  \psi^{\dagger}_{\sigma}(x)\,
V(x)
   \,\psi^{\phantom{\dagger}}_{\sigma}(x) \;,
   \nonumber \\
  {\cal H}_U &=& 
{1\over2} \sum_{\sigma \sigma '} 
\int_{0}^{L} \!\!\! \int_{0}^{L} 
  \!\! dx\, dx'  \   
 \psi^{\dagger}_{\sigma}(x)\,
  \psi^{\dagger}_{\sigma'}(x')
   \ U(x,x') \ 
   \psi^{\phantom{\dagger}}_{\sigma'}(x')
   \,\psi^{\phantom{\dagger}}_{\sigma}(x)
\;,
\nonumber 
\end{eqnarray}
where $\, \psi^{\dagger}_{\sigma}(x)$ 
creates an electron with spin $\sigma$ at position $x$. 
The spin-independent scattering potential 
$V(x)$ and the Coulomb repulsion $U(x,x')$ are restricted 
for the electrons staying in the central region; 
and $\,U(x,x')=U(x',x)$.
When the size of the central region $L$ is 
of the order of the Fermi wave-length, 
the model is reduced to an Anderson-Wolff impurity.\cite{AndersonModel,Wolff}
In this limit the perturbation theory
is valid for all values of $\,U$.
Our working hypothesis is that 
the ground state is changed 
continuously when the size $L$ is 
increased from an atomic size to larger one 
owing to the presence of the noninteracting leads.
Although it is not evident that 
the hypothesis is valid for whole region of the parameter space  
of the Hamiltonian,
which may be valid at least for a finite region in the parameter space
where $U$, $V$, and $L$ are small enough.
Unless, otherwise noted, we will be using units 
in which the Planck constant is $\hbar =1$.

We now consider the transmission and reflection coefficients 
for the system described by the Hamiltonian Eq.\ (\ref{eq:H}).
For this purpose, we introduce the retarded Green's function 
\begin{equation} 
G(x,x'; \omega +i0^+) 
 =  
 - i \int_0^{\infty} \! \! dt \,
   \left \langle \,   \left\{\,
   \psi^{\phantom{\dagger}}_{\sigma} (x,t), 
    \  \psi^{\dagger}_{ \sigma} (x',0)  
              \,  \right\}\, \right \rangle  \, e^{i\, (\omega +i0^+)\, t} \;,
   \label{eq:G_RET}  
\end{equation} 
where
$\,{\cal O}(t) \equiv 
       e^{i{\cal H}t} {\cal O} e^{- i{\cal H}t}$, 
$\langle \cdots \rangle$ denotes the thermal average
$\mbox{Tr} \left [ \, e^{-\beta  {\cal H} }\, {\cdots}
\right ] /  \mbox{Tr}  \, e^{-\beta  {\cal H} }$ 
with $\,\beta$ being the inverse temperature $1/T$, and
the curly brackets denote the anticommutator.  
The spin index has been omitted from the left-hand side
because the expectation value is assumed to be 
independent of whether spin is up or down.
The Dyson equation for the Green's function is written, in the real space, as
\begin{eqnarray} 
  G(x,x'; z)   &=& G_0(x,x'; z)
    + \int_{0}^{L} \!\!\! \int_{0}^{L}  \!\! dx_1 dx_2  \   
  G_0(x,x_1; z)\ \Sigma(x_1,x_2; z)\ G(x_2,x'; z) \;, 
  \label{eq:Dyson_S}
       \\
   &=&  G_0(x,x';z) 
+ \int_{0}^{L} \!\!\! \int_{0}^{L}  \!\! dx_1 dx_2  \   
 G_0(x,x_1; z)\ {\cal T}(x_1,x_2; z)\ G_0(x_2,x'; z) \;,
  \label{eq:Dyson_T}
\end{eqnarray} 
where $G_0(x,x'; z)$ is the free Green's function 
corresponding to ${\cal H}_0$,
and the region of the integral is taken to be $\,0 \leq x_1,\, x_2 \leq L\,$ 
because the Coulomb interaction and 
the scattering potential are restricted for electrons in the central region. 
The proper self-energy $\,\Sigma(x_1,x_2; z)$ 
and the scattering matrix ${\cal T}(x_1,x_2; z)$ 
contain both of the effects due to ${\cal H}_U$ and ${\cal H}_V$.
Note that the functions introduced here are symmetric with respect 
to the permutations of
$x$ and $x'$ because of the time reversal symmetry of $\,{\cal H}$
\begin{equation}
 \left\{ 
 \begin{array}{lll}
 G(x,x'; z)  &=& G(x',x; z) \\ 
 \Sigma(x,x'; z)  &=&  \Sigma(x',x; z)  \\
 {\cal T}(x,x'; z)  &=&  {\cal T}(x',x; z) 
 \end{array} \right.\;. 
\label{eq:symmetry}
\end{equation}
Therefore, the discontinuity of the Green's function 
at the real axis in the complex $z$-plane corresponds to 
the imaginary part 
\begin{equation}
 \left[\, G(x,x'; \varepsilon+i0^+)
- G(x,x'; \varepsilon-i0^+) \,\right] / 2i \ = \
\mbox{Im}\, G(x,x'; \varepsilon+i0^+) \;,
\label{eq:Im_G}
\end{equation}
and the similar relation holds also for 
$\Sigma(x,x'; z)$ and ${\cal T}(x,x'; z)$. 
Specifically, 
the free retarded Green's function is obtained, 
for $\,\omega > -\mu$, as 
\begin{equation}
G_0(x,x';\omega +i 0^+) =
 -i \pi \rho(\omega) \, e^{i\,k\,|x-x'|} \;,
\label{eq:G0_ret}
\end{equation}
where $\,k\equiv\sqrt{2m(\omega+\mu)}$, 
and $\,\rho(\omega) \equiv m/\pi k\,$ is the density of states.

When both $\,x$ and $\,x'$ belong to the leads,
the asymptotic form of the full Green's function 
can be written explicitly, 
making use of Eqs.\ (\ref{eq:Dyson_T}) and (\ref{eq:G0_ret}): 
\begin{equation}
G(x, x';\xi_k+i0^+) = 
 \, e^{i\,kx} 
       \left[\, 1 - i \pi \rho(\xi_k) \ {\cal T}_{kk}(\xi_k+i0^+)
       \,\right] 
   \left\{\,- i \pi \rho(\xi_k) \,\right\} 
        e^{-i\,kx'}
  \label{eq:G_AB}   
\end{equation}
for $\,x'< 0\,$ and $\,L<x\,$;  
\begin{equation}
G(x, x';\xi_k+i0^+) =  
             \left[\, e^{i\,kx} 
         - i \pi \rho(\xi_k) \ {\cal T}_{-kk}(\xi_k+i0^+) \ e^{-i\,kx}
       \,\right]  \left\{\,- i \pi \rho(\xi_k) \,\right\}e^{-i\,kx'}
  \label{eq:G_AA}   
\end{equation}
for $\,x'< x <0\,$.  Here $\,{\cal T}_{k'k}(z)$ is 
the Fourier transform of $\,{\cal T}(x,x'; z)$ 
\begin{equation}
{\cal T}_{k'k}(\omega +i0^+)\ = \ 
 \int_{0}^{L} \!\!\! \int_{0}^{L}  \!\! dx_1 dx_2  \  
e^{-i\, k'x_1}\ {\cal T}(x_1,x_2; \omega +i0^+) \ e^{i\, kx_2} \:, 
\label{eq:Tkk'}
\end{equation}
and the frequency $\omega$ is taken to be an on-shell value 
$\,\omega = \xi_k\,$ [$\equiv k^2/2m - \mu$] 
in Eqs.\ (\ref{eq:G_AB}) and (\ref{eq:G_AA}). 
Physically, the retarded Green's function $G(x, x';\xi_k+i0^+)$
represents a propagation of a single-particle-like excitation; 
an additional electron for $\xi_k >0$, 
or a hole for $-\mu<\xi_k <0$.
When $\xi_k >0$, for instance, 
the right-hand side of Eq.\ (\ref{eq:G_AB}) 
is relating to a probability amplitude for the process in which
an electron created
at $x'<0$ in the left lead is transmitted to 
the right lead with $L<x$. 
Similarly, Eq.\ (\ref{eq:G_AA}) represents 
the process in which an electron created at $x'<0$ 
propagates as an incident wave $e^{i\,kx}$ or 
 a reflected wave $e^{-i\,kx}$ for $x'<x<0$.
Therefore it seems natural to define 
the transmission and reflection coefficients using 
 Eqs.\ (\ref{eq:G_AB}) and (\ref{eq:G_AA}); 
\begin{equation} 
 \left \{\, 
  \begin{array}{lll}
  t(\xi_k) &\equiv&
        1 - i \pi \rho(\xi_k) \  {\cal T}_{kk}(\xi_k+i0^+)\\
  r(\xi_k) &\equiv&
      \ \  - i  \pi \rho(\xi_k) \ {\cal T}_{-kk}(\xi_k+i0^+)\ 
  \end{array} \right. \;.
  \label{eq:coefficients}
\end{equation} 
With this definition,
the unitarity condition $\,|t(\xi_k)|^2 + |r(\xi_k)|^2 =1\,$ 
is expressed in the form of an optical theorem
\begin{equation} 
  \mbox{Im}\, {\cal T}_{kk}(\xi_k+i0^+)
   \ = \  -\,{\pi \rho(\xi_k) \over 2} 
        \, \left[\, |{\cal T}_{kk}(\xi_k+i0^+)|^2 + 
             |{\cal T}_{-kk}(\xi_k+i0^+)|^2 \,\right] \;.
 \label{eq:optical}
\end{equation} 
In the non-interacting system,
the coefficients defined by Eq.\ (\ref{eq:coefficients}) 
coincide with the scattering coefficients 
which are obtained from the asymptotic form of the wave function,
and thus Eq.\ (\ref{eq:optical}) holds automatically.
However, 
when the Coulomb interaction is switched on,
it is not obvious whether Eq.\ (\ref{eq:optical}) holds or not.   
We show in the following that 
the unitarity condition Eq.\ (\ref{eq:optical}) 
holds at $\,T=0\,$ for the single-particle-like excitation 
at Fermi energy $\xi_k=0$
if the assumption $iii$) mentioned in the Introduction holds 
for any positions $x$ and $x'$ in the central region as 
\begin{equation}
\mbox{Im}\, \Sigma(x,x'; \,i0^+) =0 \;.
\label{eq:Im_S}
\end{equation}
For simplicity,
we temporarily write the integral equation 
Eq.\ (\ref{eq:Dyson_S}) in the operator form, 
$\, \mbox{\boldmath $G$} = \mbox{\boldmath $G_0$} + 
   \mbox{\boldmath $G_0  \Sigma \, G$}$,
and take the frequency for the operators to be $z=i 0^+$. 
In this form the ${\cal T}$-matrix is written as
\begin{equation}
\mbox{\boldmath ${\cal T}$} = 
\mbox{\boldmath $\Sigma$} +  
\mbox{\boldmath $\Sigma\, G_0 \Sigma$} +   
  \mbox{\boldmath $\Sigma\, G_0 \Sigma \, G_0 \Sigma$} + \; \cdots \;.
\label{eq:Dyson3}
\end{equation}
Since Eq.\ (\ref{eq:Im_S}) means that 
$\mbox{\boldmath $\Sigma$}$ is  Hermitian at $T=0$, i.e.  
$\mbox{\boldmath $\Sigma$} = \mbox{\boldmath $\Sigma^{\dagger}$}$, 
the optical theorem can be derived following 
essentially the same way with that in the noninteracting case 
making use of an operator identity\cite{Hewson} 
\begin{eqnarray}
\mbox{\boldmath ${\cal T}$} - 
\mbox{\boldmath ${\cal T}^{\dagger}$}
\ &=&\ 
    ( \mbox{\boldmath $\Sigma$}
 + \mbox{\boldmath $\Sigma \, G_0^{\dagger} \, \Sigma$}
 + \, \cdots \, ) 
      \  
     (\mbox{\boldmath $G_0^{\phantom{\dagger}}$} 
           - \mbox{\boldmath $G_0^{\dagger}$})   
     \  
    ( \mbox{\boldmath $\Sigma$} 
  + \mbox{\boldmath $\Sigma \, G_0^{\phantom{\dagger}} \Sigma$}
     + \, \cdots \, )
   \nonumber \\
&=& \ \mbox{\boldmath ${\cal T}^{\dagger}$}\, 
     (\mbox{\boldmath $G_0^{\phantom{\dagger}}$} 
           - \mbox{\boldmath $G_0^{\dagger}$})   
\, \mbox{\boldmath ${\cal T}$} \;.
    \label{eq:OperatorID} 
\end{eqnarray}
A diagonal element in the wave-number indices    
is written as
\begin{equation}
{\cal T}_{kk}^{\phantom{*}}(i0^+)
 - {\cal T}_{kk}^{*}(i0^+)
\ = \  - \int_{-\infty}^{+\infty} \! {dk'\over 2\pi} \ 
  {\cal T}_{k'k}^{*}(i0^+)\  
 2 \pi i  \delta(\xi_{k'}) \ {\cal T}_{k'k}(i0^+) 
\;,
    \label{eq:OperatorID_diagonal} 
\end{equation}
where $k$ is the Fermi wave number which satisfies $\xi_k=0$. 
Since the contribution of the integral 
in the right-hand side of Eq.\ (\ref{eq:OperatorID_diagonal}) 
comes only from $k'=\pm k$,
the optical theorem  Eq.\ (\ref{eq:optical}) holds 
for the single-particle-like  excitation at $\xi_k=0$.

As already mentioned, the assumption Eq.\ (\ref{eq:Im_S}) 
was introduced by Langer and Ambegaokar
to derive the Friedel sum rule 
for interacting Fermi systems:\cite{LangerAmbegaokar}
it is expressed in the present case as
\begin{eqnarray}
  \Delta n \ &\equiv& \ - {1 \over \pi}
\int_{-\infty}^{0} \!\! d \omega \!\!
\int_{-\infty}^{+\infty} \!\! dx \  
     \mbox{Im}\left [ \, G(x,x; \omega+i0^+) -  G_0(x,x; \omega+i0^+)
      \, \right ]   \nonumber \\
 &=&\  {1 \over 2\pi i} \  
       \mbox{Tr}\,  \log \left[\,\mbox{\boldmath $I$} + 
       ( \mbox{\boldmath $G_0^{\phantom{\dagger}}$}
         - \mbox{\boldmath $G_0^{\dagger}$}) \,
           \mbox{\boldmath ${\cal T}$}\,\right]  \;,
        \label{eq:Friedel}         
\end{eqnarray}
where the frequency for the operators 
in the second line is taken to be $z=i 0^+$.
We note that Eq.\ (\ref{eq:Im_S}) can be derived 
with the perturbation theory in the Coulomb interaction
and the calculation is performed in the next section.

For the later convenience, 
we introduce here the Matsubara Green's function 
\begin{equation} 
G(x,x'; i\varepsilon_n) 
 =   - \int_0^{\beta} \! \! d\tau \,
   \left \langle \, T_{\tau} \,  
   \psi^{M\phantom{\dagger}}_{\sigma} \!\!(x,\tau) 
    \  \psi^{\dagger M}_{ \sigma} (x',0)  
                  \,  \right \rangle  \, e^{i\, \varepsilon_n \tau} \;,
   \label{eq:G_Matsubara}  
\end{equation} 
where
$\,{\cal O}^M(\tau) \equiv 
       e^{\tau  {\cal H}} {\cal O} e^{- \tau  {\cal H}}$, 
and $\varepsilon_n = (2n+1)\pi/\beta$. 
The Dyson equation for the Matsubara Green's function 
is obtained by setting $z=i\varepsilon_n$ 
in Eqs.\ (\ref{eq:Dyson_S}) and (\ref{eq:Dyson_T}). 
The free Matsubara Green's function is obtained as
\begin{eqnarray} 
G_0(x,x'; i\varepsilon) 
&=&  m \, {e^{i\, \kappa(i\varepsilon) \,|x-x'|} \over 
     i\,  \kappa(i\varepsilon)}
   \label{eq:G0_Matsubara}  \;,\\
  \kappa(i\varepsilon) &\equiv& \sqrt{2m}\, 
                          \left[\,\mu^2+\varepsilon^2\,\right]^{1/4}
                             \, e^{i\,\varphi(i\varepsilon)/2}\,
                             \, \mbox{sgn} \, \varepsilon \;,
   \label{eq:kappa}  \\
   \varphi(i\varepsilon) &\equiv& \arg (\mu + i\varepsilon) \;.
   \label{eq:phi}  
\end{eqnarray} 
where $\,\mbox{sgn} \, \varepsilon\,$ is the sign function.
The range of the value of phase $\varphi(i\varepsilon)$ is defined as 
$\, -\pi  < \varphi \leq \pi$. Thus
in the limit $\varepsilon \to 0^{\pm}$; 
$\,\varphi(i0^{\pm}) =0$ for $\,\mu>0$, 
whereas $\,\varphi(i0^{\pm}) =\pm\pi$ for $\,\mu<0$.
Now we note some characteristic features 
in the limit $\varepsilon \to 0^+$.
When $\,\mu>0$,
$\,G_0(x,x'; i0^+)$ is an oscillatory function 
of $\,|x-x'|$, and 
the singular contribution with the sign function
in the right-hand side of Eq.\ (\ref{eq:G0_Matsubara}) 
corresponds to the imaginary part of the retarded function for $\omega=0$, 
i.e. $\mbox{Im}\, G_0(x,x'; i0^+)$, 
owing to the symmetry property 
Eqs.\ (\ref{eq:symmetry}) and (\ref{eq:Im_G}).
On the other hand, when $\,\mu<0$, 
$\,G_0(x,x'; i0^+)$ is a real and exponentially decreasing function 
of $\,|x-x'|$, 
and the singular term disappears from
the right-hand side of Eq.\ (\ref{eq:G0_Matsubara}) 
because $\, \exp[i\varphi(i0^{\pm})/2]\, \mbox{sgn} \, 0^{\pm} = i$.

\section{IMAGINARY PART OF THE SELF-ENERGY}
\label{sec:self-energy}

In this section, we calculate the imaginary part of the proper self-energy
up to the order $\varepsilon^2$ 
with the perturbation theory in 
the Coulomb interaction.\cite{Yamada4,AGD,YY_PAM}
We show that the imaginary part 
is proportional to $\varepsilon^2$ at $\,T=0$, 
and thus Eq.\ (\ref{eq:Im_S}) holds in the perturbation theory.

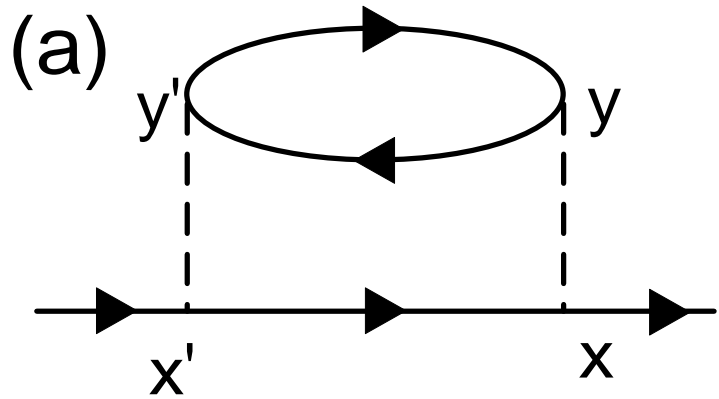
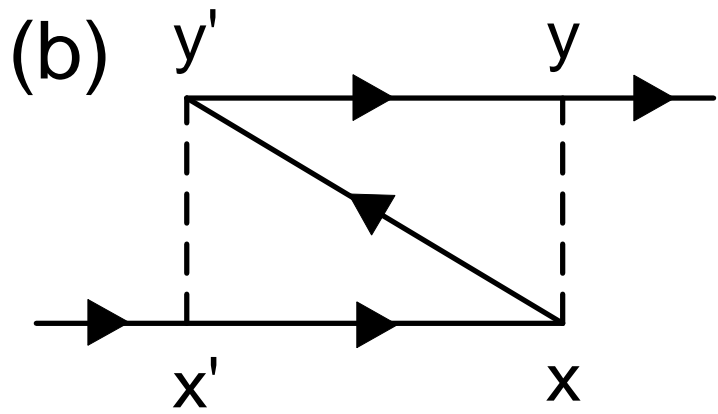
\begin{figure}
\caption{ Second order skeleton diagrams 
for the proper  self-energy
$\,\Sigma(x, x';i\varepsilon_n)$.}
\label{fig:self2}
\end{figure}

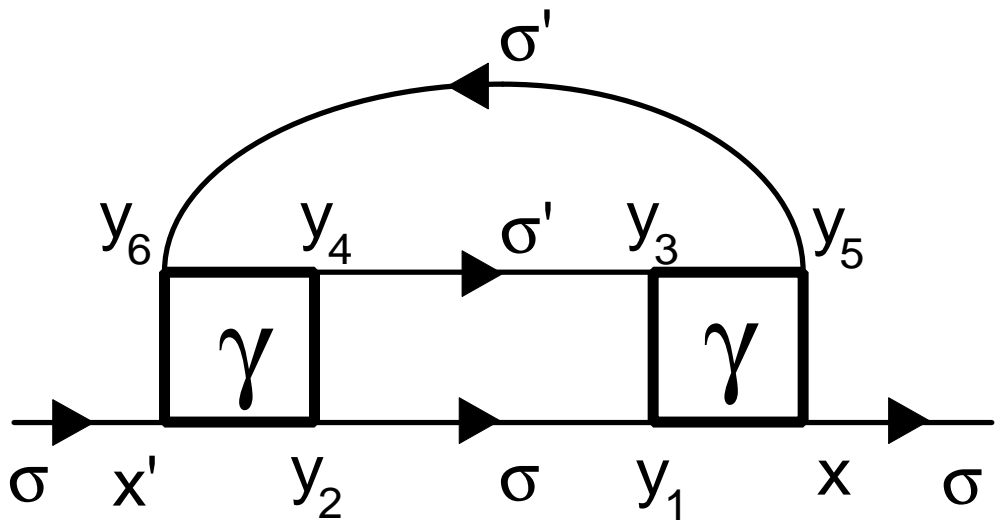
\begin{figure}
\caption{ Second order skeleton diagrams 
in terms of the bare vertex function 
$\,\gamma_{\sigma\sigma'}(x_4, x_3; x_2, x_1)$.} 
\label{fig:self2_s}
\end{figure}

As already mentioned, the self-energy 
satisfies the relation corresponding to Eq.\ (\ref{eq:Im_G}), 
\begin{equation}
\mbox{Im}\, \Sigma(x,x'; \varepsilon+i0^+) = 
  \left[\, \Sigma(x,x'; \varepsilon+i0^+) 
 - \Sigma(x,x'; \varepsilon-i0^+) \,\right] / 2i \:,
\label{eq:Im_S_dis}
\end{equation}
owing to the symmetry property Eq.\ (\ref{eq:symmetry}). 
It means that the imaginary part of the retarded self-energy 
$\Sigma(x,x'; \varepsilon+i0^+)$ 
is related to the singular 
$\,\mbox{sgn} \, \varepsilon_n\,$ 
contributions 
in the Matsubara function $\Sigma(x,x'; i\varepsilon_n)$ 
by the analytic continuation, 
and this correspondence holds for any $x$ and $x'$.
Therefore we can obtain the $\varepsilon^2$-term  
following the calculations 
in the case of the usual Fermi liquid\cite{Yamada4,AGD,YY_PAM} 
in which the self-energy is a diagonal quantity.
Yamada showed precisely that 
the singular contribution up to the order $\varepsilon^2_n$ 
comes only from diagrams which contain 
three intermediate Green's functions.\cite{Yamada4,Hewson}
Fig.\ \ref{fig:self2} shows  
the skeleton diagram for the simplest example,
and the expression for the imaginary part is written 
after the analytic continuation as\cite{AGD} 
\begin{eqnarray}
& &
\!\!\!\!\!\!\!\!
\mbox{Im}\, \Sigma^{(2)}(x,x'; \,\varepsilon+i0^+) 
\nonumber
\\ 
 &=&\ 
  \int_{-\infty}^{+\infty} \! {d\varepsilon' d\varepsilon'' \over (2 \pi)^2}  
  \,\left[\, \coth{\varepsilon'-\varepsilon \over 2T} 
  - \tanh{\varepsilon' \over 2T} 
  \,\right]
  \left[\, \tanh{\varepsilon'' \over 2T} 
  - \tanh{\varepsilon''+\varepsilon-\varepsilon' \over 2T} 
  \,\right]  
\nonumber\\
 & &\ 
\times 
 \int_{0}^{L} \!\!\! \int_{0}^{L} \!\!\! 
dy dy' \,
U(x', y')\, U(x, y) \  
    \mbox{Im}\, G(y',y; \varepsilon''+i0^+) \,  
\nonumber\\
 & &\  
\times 
\biggl\{\   
  2\  \mbox{Im}\, G(x,x'; \varepsilon'+i0^+) \,
    \mbox{Im}\, G(y,y'; \varepsilon+\varepsilon''-\varepsilon'+i0^+) 
\nonumber\\
 & & \qquad   
     -\  \mbox{Im}\, G(y,x'; \varepsilon'+i0^+) \,
    \mbox{Im}\, G(x,y'; \varepsilon+\varepsilon''-\varepsilon'+i0^+) 
\ \biggr\}
\;. 
\label{eq:self2}
\end{eqnarray}
Here $G$ is the full Green's function which contains all the effects 
of the scattering potential and the Coulomb interaction.
At $\,T=0$, the hyperbolic functions in Eq.\ (\ref{eq:self2}) 
are replaced by the sign functions as
  $\,\left[\, \mbox{sgn}(\varepsilon'-\varepsilon ) 
  - \mbox{sgn}(\varepsilon' ) 
  \,\right]
  \left[\, \mbox{sgn}(\varepsilon'' ) 
  - \mbox{sgn}(\varepsilon''+\varepsilon-\varepsilon' )
  \,\right]$.
Thus the region of the integral for the frequencies are reduced to 
$\displaystyle\int_0^{\varepsilon}\!\!d\varepsilon' 
\!\! \displaystyle\int_{\varepsilon'-\varepsilon}^0\!\!d\varepsilon''$, 
and the area of the integral region becomes $\varepsilon^2/2$.  
Therefore the expression for small $\varepsilon$ is obtained  
setting the frequencies for the Green's functions in Eq.\ (\ref{eq:self2}) 
to be zero; 
\begin{eqnarray}
& &
\!\!\!\!\!\!\!\!
\mbox{Im}\, \Sigma^{(2)}(x,x'; \,\varepsilon+i0^+) 
\nonumber
\\ 
 &\simeq& {\varepsilon^2 \over 2 \pi^2}
 \int_{0}^{L} \!\!\! \int_{0}^{L} \!\!\! dy dy' \,
U(x', y')\, U(x,y)\   
    \mbox{Im}\, G(y',y; i0^+) \,  
\nonumber\\
 & &  
\times 
   \left\{\, 2\ \mbox{Im}\, G(x,x'; i0^+) \,
    \mbox{Im}\, G(y,y'; i0^+) 
     -  \mbox{Im}\, G(y,x'; i0^+) \,
    \mbox{Im}\, G(x,y'; i0^+) \,\right\} \;. 
\label{eq:self2a}       
\end{eqnarray}
The imaginary part is proportional to $\varepsilon^2\,$ 
for any  $x$ and $x'$ 
because the $\varepsilon^2$-dependence 
is coming from the hyperbolic functions. 
It is convenient for proceeding next step 
to rewrite Eq.\ (\ref{eq:self2a}) by introducing the bare vertecies 
$\gamma_{\uparrow \downarrow}$ and $\gamma_{\uparrow \uparrow}$ 
[see Fig.\ \ref{fig:self2_s}]
\begin{eqnarray}
& &
\!\!\!\!\!\!\!\!
\mbox{Im}\, \Sigma^{(2)}(x,x'; \,\varepsilon+i0^+) 
\nonumber
\\ 
 &\simeq& {\varepsilon^2 \over 2 \pi^2}
 \int_{0}^{L} \!\!\!\!\! \cdots \!\! \int_{0}^{L} 
   \prod_{i=1}^6 dy_i \ 
    \mbox{Im}\, G(y_6,y_5; i0^+) \,  
    \mbox{Im}\, G(y_1,y_2; i0^+) \,  
    \mbox{Im}\, G(y_3,y_4; i0^+) \,  
\nonumber\\
 & &  
\times 
   \left[\, 
         \gamma_{\uparrow \downarrow}(x, y_5; y_3,y_1)\,
         \gamma_{\uparrow \downarrow}(y_2, y_4; y_6, x')
       + {1 \over 2}\, 
         \gamma_{\uparrow \uparrow}(x, y_5; y_3, y_1)\,
         \gamma_{\uparrow \uparrow}(y_2, y_4; y_6, x')
      \,\right] \;,
\label{eq:self2b}
\end{eqnarray}
where 
\begin{eqnarray}
\gamma_{\uparrow \downarrow}(x_4,x_3; x_2, x_1) 
&\equiv& U(x_1,x_2)\, \delta(x_1-x_4)\, \delta(x_2-x_3) \;,  \\
\gamma_{\uparrow \uparrow}(x_4,x_3; x_2, x_1) 
&\equiv& U(x_1,x_2) 
\left[\, \delta(x_1-x_4)\, \delta(x_2-x_3)
 \,-\, \delta(x_1-x_3)\, \delta(x_2-x_4) \,\right]\;.
\end{eqnarray}

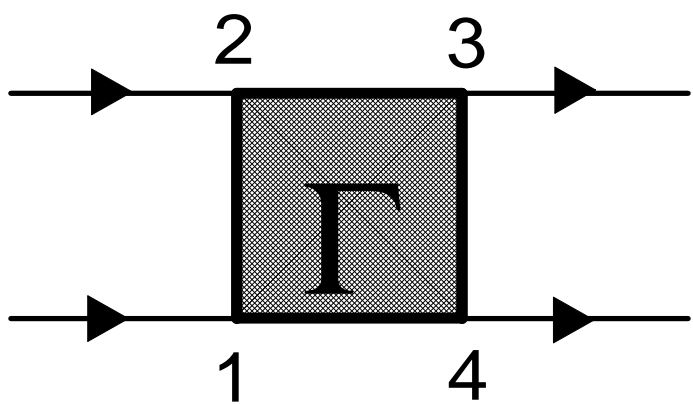
\begin{figure}
\caption{ Full vertex function
$\,\Gamma_{\sigma_4 \sigma_3; \sigma_2 \sigma_1} 
(x_4, x_3; x_2, x_1: 
i\varepsilon_4, i\varepsilon_3; i\varepsilon_2,i\varepsilon_1)$ .}
\label{fig:vertex}
\end{figure}

We now include the higher order terms 
following Yamada,\cite{Yamada4,YY_PAM}
which can be completed just by replacing
the bare vertices $\gamma_{\sigma \sigma'}$ 
with the full ones  $\Gamma_{\sigma \sigma'}$ [see Fig.\ \ref{fig:vertex}]
\begin{equation} 
         \Gamma_{\sigma \sigma '}(x_4, x_3 ; x_2, x_1)
         \equiv
\Gamma_{\sigma \sigma '; \sigma' \sigma }
(x_4, x_3; x_2, x_1: 0, 0; 0, 0) \;.
\end{equation} 
This is because 
the singular $\varepsilon_n^2 \mbox{sgn}\, \varepsilon_n$ contributions 
in the Matsubara function $\Sigma(x,x'; i\varepsilon_n)$ 
can be obtained by summing up all the diagrams 
with the three intermediate Green's functions. 
Therefore,
we can write down the expression 
which is exact up to $\varepsilon^2$-term at $\,T=0\,$ 
by symmetrizing Eq.\ (\ref{eq:self2b}) with respect to 
the interchange of $x$ and $x'$ making use of Eq.\ (\ref{eq:symmetry}),
then adding the equivalent expression in which the set of variables  
for the integration $\{y_5,y_3,y_1\}$ and $\{y_6,y_4,y_2\}$ are interchanged,
and finally replacing the bare vertices by the full ones;  
\begin{eqnarray}
& &
\!\!\!\!\!\!\!\!
\mbox{Im}\, \Sigma(x,x'; \,\varepsilon+i0^+) 
\nonumber
\\ 
 &\simeq& {\varepsilon^2 \over 8 \pi^2}
 \int_{0}^{L} \!\!\!\!\! \cdots \!\! \int_{0}^{L} 
   \prod_{i=1}^6 dy_i \ 
    \mbox{Im}\, G(y_6,y_5; i0^+) \,  
    \mbox{Im}\, G(y_1,y_2; i0^+) \,  
    \mbox{Im}\, G(y_3,y_4; i0^+) \,  
\nonumber\\
 & &  
\times 
   \biggl[\, 
         \Gamma_{\uparrow \downarrow}(x',y_6;y_4, y_2) \,
         \Gamma_{\uparrow \downarrow}(y_1,y_3;y_5, x)
       +  \Gamma_{\uparrow \downarrow}(x,y_5;y_3, y_1) \,
         \Gamma_{\uparrow \downarrow}(y_2,y_4;y_6, x')
\nonumber\\
 & &  \ \ 
       +  \Gamma_{\uparrow \downarrow}(x',y_5;y_3, y_1) \,
         \Gamma_{\uparrow \downarrow}(y_2,y_4;y_6, x)
      +  \Gamma_{\uparrow \downarrow}(x,y_6;y_4, y_2) \,
         \Gamma_{\uparrow \downarrow}(y_1,y_3;y_5, x')
\nonumber\\
 & & \ \  
       + {1 \over 2}\, \Gamma_{\uparrow \uparrow}(x',y_6;y_4, y_2) \,
         \Gamma_{\uparrow \uparrow}(y_1,y_3;y_5, x)
       + {1 \over 2}\, \Gamma_{\uparrow \uparrow}(x,y_5;y_3, y_1) \,
         \Gamma_{\uparrow \uparrow}(y_2,y_4;y_6, x')
\nonumber\\
 & & \ \  
       + {1 \over 2}\, \Gamma_{\uparrow \uparrow}(x',y_5;y_3, y_1) \,
         \Gamma_{\uparrow \uparrow}(y_2,y_4;y_6, x) 
       + {1 \over 2}\, \Gamma_{\uparrow \uparrow}(x,y_6;y_4, y_2) \,
         \Gamma_{\uparrow \uparrow}(y_1,y_3;y_5, x') 
      \,\biggr] \;.
\label{eq:self_all}
\end{eqnarray}
Here we note that the full vertex function 
$\Gamma_{\sigma_4 \sigma_3; \sigma_2 \sigma_1 }
(x_4, x_3; x_2, x_1: 
i\varepsilon_4, i\varepsilon_3; i\varepsilon_2,i\varepsilon_1)$ 
is antisymmetric 
with respect to interchanges 
of the arguments (together with the spin suffixes)
in the first or second pair: $1$ and $2$, or $3$ and $4$.\cite{Lifshitz}
In the symmetrized expression, Eq.\ (\ref{eq:self_all}),
the eight terms in the bracket consist of four pairs 
each of which is connected 
by the symmetry operation for the time reversal; 
$(x_4,x_3;x_2, x_1) 
\Leftrightarrow (x_1,x_2;x_3, x_4)$.
Consequently, the imaginary part of the proper self-energy is proportional 
to $\varepsilon^2$ and Eq.\ (\ref{eq:Im_S}) holds at $\,T=0\,$
as long as the integral in Eq.\ (\ref{eq:self_all}) is finite, i.e.,
in the case the perturbation theory is valid.

\section{CONDUCTANCE}
\label{sec:Kubo}

In this section, we consider the expression of the dc conductance $\,g\,$ 
applying the perturbation analysis\cite{Shiba} 
to the current-current correlation function.
We show that the contributions of the vertex correction 
to $\,g$ is zero at $T=0$ when the currents are measured 
in the noninteracting leads. 
The resultant expression,  
Eq.\ (\ref{eq:cond2}), is written 
in terms of the transmission coefficient 
for the single-particle-like excitation at the Fermi energy, $t(0)$, 
defined by Eq.\ (\ref{eq:coefficients}).

When the electric field is applied to the central region,
the dc conductance within the Kubo formalism is given by
\begin{equation} g \ =   \ \lim_{\omega \to 0}\, 
 { K(x,x';\omega+i0^+ ) - K(x,x';i0^+) \over  i \omega} \;,
\label{eq:Kubo} 
\end{equation}
where
\begin{eqnarray}
& & K(x,x';\omega + i0^+)\ = \  
i \int_0^{\infty} \! \! dt \,
\langle\,  \left[\, j(x, t)\,, j(x',0) \,\right]\, \rangle 
\ e^{i\,(\omega +i0^+)\, t } 
\;, \label{eq:K_R}  \\
& & \quad  
 j(x)\  =\   \sum_{\sigma} {e \over 2 m i} 
  \left[\,
 \psi^{\dagger}_{\sigma}(x)\ {\partial \psi_{\sigma}(x) \over \partial x} 
\ - \
  {\partial \psi^{\dagger}_{\sigma}(x) \over \partial x}\ \psi_{\sigma}(x)
\,\right]  \;. 
\end{eqnarray}
As discussed by Fisher \& Lee,\cite{Fisher,Lee} 
$\,x$ and $\,x'$ are arbitrary in Eq.\ (\ref{eq:Kubo}) 
and $\,g\,$ is independent of the choice of the positions 
$x$ and $x'$ owing to the current conservation specific to one dimension. 
This feature can be seen explicitly 
in the Lehmann representation
\begin{equation}
 g = {\pi \beta \over  Z }
    \,  \sum_{\alpha \alpha '} \, e^{-\beta E_{\alpha}}
              \langle \alpha |j(x)|\alpha ' \rangle 
              \langle \alpha ' |j(x')|\alpha  \rangle \, 
              \delta(E_{\alpha} - E_{\alpha '}) \;,
\label{eq:Lehmann}
\end{equation}
where $\,Z \equiv \mbox{Tr}\, e^{-\beta {\cal H}}$, 
$\, |\alpha \rangle$ is the eigenstate of $\,{\cal H}$, 
and $\,E_{\alpha}$ is its eigenvalue.
For obtaining this representation,
we have used the relation 
caused by the time reversal symmetry: 
$\,K(x,x'; z) = K(x',x;z)\,$ and 
\begin{equation}
 \left[\, K(x,x'; \omega+i0^+)
- K(x,x'; \omega-i0^+) \,\right] / 2i \ = \
\mbox{Im}\, K(x,x'; \omega+i0^+) \;.
\label{eq:Im_K}
\end{equation}
In Eq.\ (\ref{eq:Lehmann}), 
the matrix element $\,\langle \alpha |j(x)|\alpha ' \rangle\,$
is independent of $x$ because $E_{\alpha}=E_{\alpha '}$.\cite{Fisher} 
This can be confirmed by multiplying the equation of continuity 
$\partial \rho(x,t) / \partial t \, +  \partial j(x,t)/ \partial x= 0$ 
 by $\,\langle \alpha |$ on the left 
and by $\,|\alpha' \rangle$ on the right, where 
$\,\rho \equiv e \sum_{\sigma} 
\psi^{\dagger}_{\sigma}\,\psi^{\phantom{\dagger}}_{\sigma}\,$.

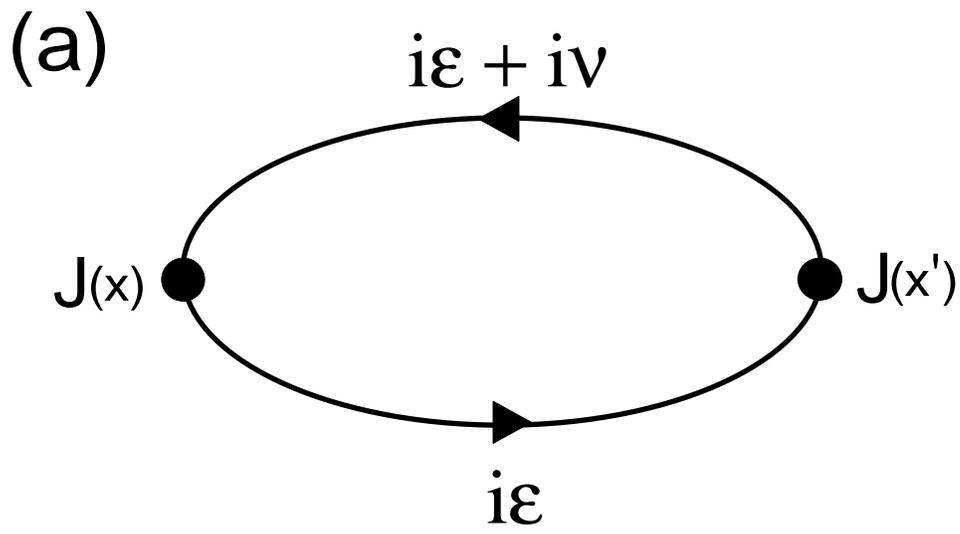
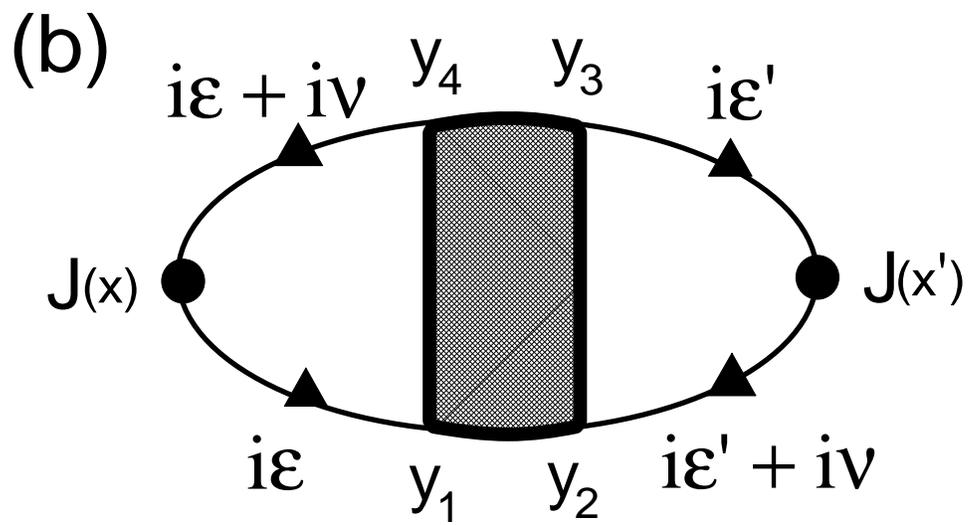
\begin{figure}
\caption{ Current-current correlation function $\,K(x, x'; i\nu)$.}
\label{fig:K}
\end{figure}

In the Kubo formalism, Eq.\ (\ref{eq:Kubo}), 
the dc conductance is obtained from  
the $\omega$-linear imaginary part of $\,K(x,x';\omega+ i0^+ )$.
In addition, Eq.\ (\ref{eq:Im_K}) means that 
the imaginary part corresponds to the discontinuity of $\,K(x,x';z)\,$ 
at the real axis in the complex $z$-plane, i.e., 
the singular contribution of the form $\,\mbox{sgn} (\mbox{Im}\, z)$. 
This type of singular contribution in the correlation function 
can be extracted using the perturbation analysis 
which was originally used by Shiba to derive 
the Korringa relation in Kondo alloys.\cite{Shiba} 
In the previous study,
we applied this method to the conductance through 
a single Anderson-Wolff impurity.\cite{ao3}
In order to carry out the analysis, 
we introduce the Matsubara function 
\begin{equation} 
K(x,x';i \nu_n)\ = \  
  \int_0^{\beta} \! \! d\tau \,
\langle\, T_{\tau}\, j^M(x, \tau)\, j^M(x',0) \, \rangle \ e^{i\, \nu_n \tau} 
\;, 
\label{eq:K_M}  
\end{equation} 
where $\,\nu_n = 2 \pi n/\beta$. 
The $\omega$-linear imaginary part of $\,K(x,x';\omega+ i0^+ )$
is obtained from the
$\, |\nu|$-linear contribution of $\,K(x,x';i\nu)$
because $\,|\nu| =\, \nu\ \mbox{sgn}\, \nu \,$ 
is replaced by $\,-i \omega\,$ 
by the analytic continuation $\,i\nu \to \omega + i0^+$. 
The Matsubara function $K(x,x';i \nu)$ can be 
represented by a sum of the diagrams shown in Fig.\ \ref{fig:K}, 
and the corresponding expression for $\,T=0\,$ is written, 
replacing sums over discrete Matsubara frequencies by integrals, as    
\begin{eqnarray}
K(x,x';i \nu) \  &=& \ K^{(a)}(x,x';i \nu) \ + \ K^{(b)}(x,x';i \nu) \quad , 
\label{eq:K_nu} \\
K^{(a)}(x,x';i \nu) &=& 
- \left({e \over 2 m i}\right)^2 
\left\{ {\partial / \partial x_4}\,
 - \,{\partial / \partial x_1} \right\} 
\left\{ {\partial / \partial x_3}\, 
 - \,{\partial / \partial x_2} \right\}
\nonumber  \\
& &
\times  
\sum_{\sigma} 
\int_{-\infty}^{+\infty} \! {d\varepsilon \over 2 \pi} \ 
\biggl.
G(x_3,x_1;i\varepsilon) \,  G(x_4,x_2;i\varepsilon + i\nu)\,
\biggr|_{\scriptstyle x_1, x_4 \to x \atop \scriptstyle \ x_2, x_3 \to x'} 
\;\;,  
\label{eq:K_nu_a} \\ 
K^{(b)}(x,x';i \nu) &=& 
- \left({e \over 2 m i}\right)^2 
\left\{ {\partial/\partial x_4}\,
 - \,{\partial/\partial x_1} \right\} 
\left\{ {\partial/\partial x_3}\, 
 - \,{\partial/\partial x_2} \right\}
\nonumber  \\
& & \times \sum_{\sigma \sigma '} 
\int_{-\infty}^{+\infty} \! {d\varepsilon d\varepsilon' \over (2 \pi)^2}  
 \int_{0}^{L} \!\!\!\!\! \cdots \!\! \int_{0}^{L} 
   \prod_{i=1}^4 dy_i \, 
 G(y_1,x_1;i\varepsilon) \,  G(x_4,y_4;i\varepsilon + i\nu)\,
\nonumber  \\
& & \times 
\Gamma_{\sigma \sigma '; \sigma' \sigma}
(y_4, y_3; y_2, y_1: 
i \varepsilon + i \nu, i \varepsilon';
i \varepsilon' + i \nu, i \varepsilon) 
\nonumber  \\
& & \times
\biggl.
G(x_3,y_3;i\varepsilon') \,  G(y_2,x_2;i\varepsilon' + i\nu)\,
\biggr|_{\scriptstyle x_1, x_4 \to x \atop \scriptstyle \ x_2, x_3 \to x'} 
\;\;.  
\label{eq:K_nu_b}
\end{eqnarray}
In what follows we choose $\,x$ to be 
in the right lead and $\, x'$ to be 
in the left lead, and thus  $x'<0<y_i<L<x$ with $i=1,\ldots,4$. 
Then the asymptotic form of the Matsubara Green's functions in 
Eqs.\ (\ref{eq:K_nu_a}) and (\ref{eq:K_nu_b})
can be written, 
making use of Eqs.\ (\ref{eq:Dyson_T}) 
and (\ref{eq:G0_Matsubara})--(\ref{eq:phi}), as
$\,G(x, y_i; i\varepsilon)  
\propto \exp[{i \kappa(i\varepsilon) \,(x-y_i)}]\, $
and $\,G(x', y_i ; i\varepsilon) 
\propto  \exp[{i \kappa(i\varepsilon) \,(y_i-x')}]\,$.
Therefore the derivative with respect to $\,x_1,\ldots,x_4\,$ 
can be done explicitly,   
and the expression valid for $x'<0$ and $L<x$ is obtained as 
\begin{eqnarray}
K^{(a)}(x,x';i \nu) &=& 
- \left({e \over 2 m }\right)^2 
\sum_{\sigma} 
\int_{-\infty}^{+\infty} \! {d\varepsilon \over 2 \pi} \, 
\{ \kappa(i\varepsilon + i\nu) - \kappa(i\varepsilon) \}^2\,
G(x',x;i\varepsilon) \,  G(x,x';i\varepsilon + i\nu) \;,
\label{eq:bubble} \\
K^{(b)}(x,x';i \nu) &=& 
- \left({e \over 2 m }\right)^2 
\sum_{\sigma \sigma '} 
\int_{-\infty}^{+\infty} \! {d\varepsilon d\varepsilon' \over (2 \pi)^2}  
 \int_{0}^{L} \!\!\!\!\! \cdots \!\! \int_{0}^{L} 
   \prod_{i=1}^4 dy_i \, 
\nonumber \\
& & \times 
\{ \kappa(i\varepsilon + i\nu) - \kappa(i\varepsilon) \}\,
G(y_1,x;i\varepsilon) \,  G(x,y_4;i\varepsilon + i\nu)\,
\nonumber\\
& &\times 
\Gamma_{\sigma \sigma '; \sigma' \sigma}
(y_4, y_3; y_2, y_1: 
i \varepsilon + i \nu, i \varepsilon';
i \varepsilon' + i \nu, i \varepsilon) 
\nonumber  \\
& & \times
\{ \kappa(i\varepsilon' + i\nu) - \kappa(i\varepsilon') \}\,
G(x',y_2;i\varepsilon') \,  G(y_3,x';i\varepsilon' + i\nu)\,
\;\;.  
\label{eq:vertex}
\end{eqnarray}

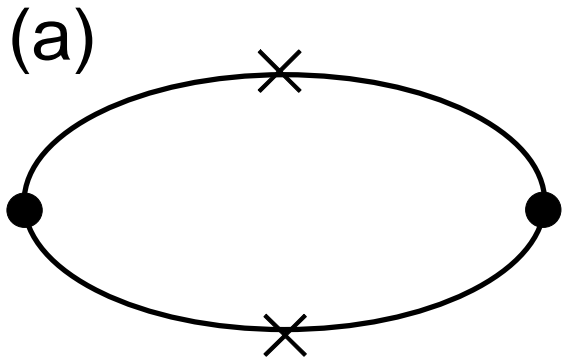
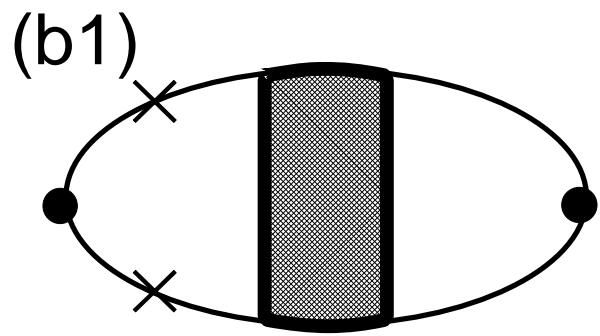
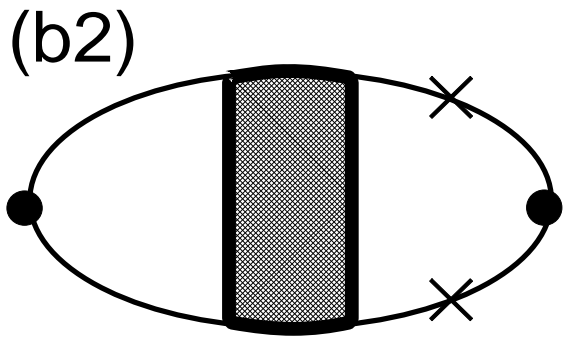
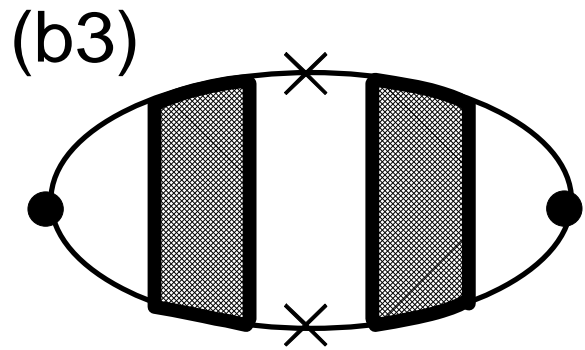
\begin{figure}
\caption{ Diagrams for the $\nu\, \mbox{sgn}\ \nu$ contribution 
in $\,K(x, x';i\nu)$.  
The two intermediate Green's functions
which yield the singular contribution are
marked with the cross.}
\label{fig:G}
\end{figure}

We can now extract the singular 
$\,\nu\, \mbox{sgn}\, \nu\,$ contribution  
from Eqs.\ (\ref{eq:bubble}) and (\ref{eq:vertex}). 
Shiba has shown precisely
that this type of singular contribution 
comes only from the diagrams 
with two intermediate Green's functions 
which carry the same frequencies in the limit $\nu \to 0$.\cite{Shiba,Hewson} 
The simplest example is the diagram Fig.\ \ref{fig:K} (a)  
in which the frequencies are assigned as 
$G(i\varepsilon)\,G(i\varepsilon+i\nu)$.
The $\,\nu\, \mbox{sgn}\, \nu\,$ contribution is originated from 
the $\,\mbox{sgn}\,\varepsilon\,$ term in 
the Matsubara Green's function: 
for instance, see Eq.\ (\ref{eq:kappa}) 
for the free Green's function.
The full Green's function can be expressed, 
by separating the singular and regular parts, as
\begin{equation}
G(x,x';i\varepsilon) = G^{'}(x,x';i\varepsilon) 
              + i\, G^{''}(x,x';i\varepsilon)\, \mbox{sgn}\, \varepsilon 
\;.
\label{eq:separate}
\end{equation}
Note that, when $\,\varepsilon=0$, 
 $G^{'}(x,x';0)$ and  $G^{''}(x,x';0)$ 
correspond to the real and imaginary parts 
of the retarded Green's function $G(x,x';i0^+)$
owing to the symmetry property Eq.\ (\ref{eq:Im_G}).
The $\,\nu\, \mbox{sgn}\, \nu$ contribution of $K(x,x';i \nu)$ 
due to the two intermediate Green's functions 
appears through the integration of the product  
$\,\mbox{sgn}(\varepsilon + \nu)\,\mbox{sgn}(\varepsilon)$ 
over $\varepsilon$ ; 
\begin{equation}
\left.
 \int_{-\infty}^{+\infty} \!\!\! d\varepsilon  \,  
 A(i \varepsilon)\  \mbox{sgn}(\varepsilon + \nu) \  
               \mbox{sgn}(\varepsilon) \,\right|_{\nu \to 0}
 = \int_{-\infty}^{+\infty} \!\!\! d\varepsilon \,  A(i \varepsilon)
    \ - \ 2\, A(0)\  \nu \ \mbox{sgn}\, \nu  \;.
\label{eq:example}
\end{equation}
Here $A(i \varepsilon)$ is a smooth function which 
can be expressed in terms of $\,G^{'}$, $G^{''}$, and so on,
and the second term in the right-hand side 
appears as a result of the derivative 
$d\, \mbox{sgn}(\varepsilon+\nu) /d\nu = 2\,\delta(\varepsilon+\nu)$.
Thus, the next thing we should do is to calculate the coefficient $A(0)$ 
taking into account the vertex correction.
In the perturbation theory, 
the diagrams which 
contain the two intermediate Green's functions 
can be classified into four groups[see Fig.\ \ref{fig:G}].\cite{Shiba} 
In the figure, 
the two intermediate Green's functions 
which yield the singular contribution 
are marked with the cross, 
and the external frequency $\,\nu$ is taken to be zero 
in the remaining part of each diagrams. 
One remarkable feature is that 
the contributions of the diagrams (b1)--(b3) become zero,
i.e., the coefficient $A(0)$ corresponding 
to these diagrams is zero. 
This is because the each of diagrams (b1)--(b3) contains 
at least one {\em current vertex\/}  
$\,\left\{ \kappa(i\varepsilon' + i\nu) - \kappa(i\varepsilon')  \right\}$, 
for which $\nu$ can be simply taken to be zero.
In contrast to (b1)--(b3), 
the contribution of the diagram (a) is finite 
because  the two intermediate Green's functions
and the {\em current vertex\/} carry the same the internal frequency 
$\varepsilon$, 
and the product $\,\mbox{sgn}(\varepsilon + \nu)\,\mbox{sgn}(\varepsilon)$ 
appears also from the square of the {\em current vertex\/} 
[see Eq.\ (\ref{eq:bubble})].
Therefore, the finite contribution of 
the $\,\nu\,\mbox{sgn}\, \nu$ term comes only from
the diagram (a), which can be calculated by substituting 
Eqs.\ (\ref{eq:kappa}), (\ref{eq:phi}) and (\ref{eq:separate}) 
into Eq.\ (\ref{eq:bubble}).
Consequently,
the dc conductance at $\,T=0\,$ 
is obtained, by reinserting $\,\hbar$, as 
\begin{equation}
 g  =  {e^2 \over \pi\hbar} \  
            {1 \over \left[\,\pi\rho(0)\,\right]^2}
            \left|\, G(x,x';i0^+)\, \right|^2  
 \label{eq:cond1}   \;,
\end{equation}
where $x$ and $x'$ is taken to be $x'<0$ and $L<x$ as already noted.
This expression can be rewritten in 
terms of the transmission coefficient $t(0)$ defined by 
Eqs.\ (\ref{eq:G_AB}) and (\ref{eq:coefficients});
\begin{equation}
 g =  {2  e^2 \over h} \  \left|t(0) \right|^2  \;.
\label{eq:cond2}
\end{equation}
These results  Eqs.\ (\ref{eq:cond1}) and (\ref{eq:cond2}) 
are generalized to a quasi-one-dimensional system 
with a number of channels in the next section.
Furthermore,
the present analysis can be applied to 
a tight-binding model and 
the results can also be generalized 
to the system on the lattice.\cite{ao5}

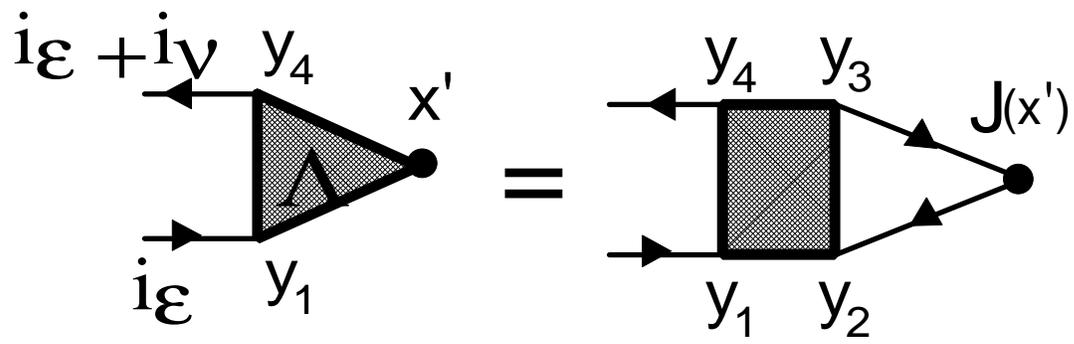
\begin{figure}
\caption{ 
Three point vertex function for the current 
$\,\Lambda(x'; y_4, y_1; i\varepsilon+i \nu, i\varepsilon)$. 
}
\label{fig:currentV}
\end{figure}

Before closing this section we consider briefly the reason 
why the contributions of the vertex correction disappear, 
and it seems to be understood in relation to the back flow 
in the leads. 
In the usual Fermi liquid theory 
the back flow can be represented in terms of  
the three point vertex function for the current,\cite{YY_PAM,Eliashberg} 
which is one parts of Eq.\ (\ref{eq:K_nu_b}) 
and can be written as 
[see Fig.\ \ref{fig:currentV}]
\begin{eqnarray}
\Lambda(x'; y_4,y_1 ; i\varepsilon+i \nu, i\varepsilon) 
&=&   
 \sum_{\sigma'} 
\int_{-\infty}^{+\infty} \! {d\varepsilon' \over 2 \pi}  
\int_{0}^{L} \!\!\! \int_{0}^{L} \!\!\!
dy_2 dy_3 \ 
\Gamma_{\sigma \sigma '; \sigma' \sigma}
(y_4, y_3; y_2, y_1: i \varepsilon + i \nu, i \varepsilon';
i \varepsilon' + i \nu, i \varepsilon) 
\nonumber  \\
& & \times
{e \over 2 m i}
\left\{ {\partial/\partial x_3}\, 
 - \,{\partial/\partial x_2} \right\}
G(x_3,y_3;i\varepsilon') \,  G(y_2,x_2;i\varepsilon' + i\nu)\,
\biggr|_{x_2, x_3 \to x'} 
\;.  
\label{eq:lambda}
\end{eqnarray}
Specifically, the value $\,\Lambda(x';y_4, y_1; 0, 0)\,$ for zero frequencies
can be related to the renormalization of the current 
caused by the back flow.\cite{YY_PAM,Eliashberg}
Since we are considering the system without the translational invariance, 
a more precise definition of the back flow should be necessary 
for a general argument. 
Nevertheless, we can show the value $\Lambda(x'; y_4, y_1; 0, 0)$ to be zero 
when $\,x'$ belongs to the noninteracting leads.
It can be shown by performing the derivative with respect 
to $x_2,\,x_3$ as it was done for obtaining Eq.\ (\ref{eq:vertex}); 
\begin{eqnarray}
\Lambda(x';y_4,y_1; i\varepsilon + i\nu, i \varepsilon) 
&=&   \pm\,
\sum_{\sigma'} 
\int_{-\infty}^{+\infty} \! {d\varepsilon' \over 2 \pi}  
\int_{0}^{L} \!\!\! \int_{0}^{L} \!\!\!
dy_2 dy_3 \ 
\Gamma_{\sigma \sigma '; \sigma' \sigma}
(y_4, y_3; y_2, y_1: i \varepsilon + i \nu, i \varepsilon';
i \varepsilon' + i \nu, i \varepsilon) 
\nonumber  \\
& & \times 
{e \over 2 m }
\{ \kappa(i\varepsilon' + i\nu) - \kappa(i\varepsilon') \}\,
    G(y_2,x';i\varepsilon' + i\nu)\, G(x',y_3;i\varepsilon') \;,
\label{eq:lambda2}
\end{eqnarray}
where $\pm$ is chosen to be $\,+$ for $\,x'<0\,$, 
and $\,-$ for $\,L<x'$. 
Since the {\em current vertex\/} 
$\{ \kappa(i\varepsilon' + i\nu) - \kappa(i\varepsilon') \}$
is zero for $\,\nu=0$, the three point function becomes 
$\,\Lambda(x'; y_4, y_1; 0, 0)=0$. 
In this sense, 
the current is not renormalized 
and the effect due to the back flow 
is absent in the noninteracting leads. 
This is caused by the fact that the asymptotic form of the Green's function 
in the leads is given by a simple superposition of the incident, 
transmitted, and reflected waves 
[see Eqs.\ (\ref{eq:G_AB}) and (\ref{eq:G_AA})].

\section{EXTENSION TO HIGHER DIMENSIONS}
\label{sec:extension}

 So far, we have considered the system in one dimension.
In this section, we generalize the results obtained in the above 
to the quasi-one-dimensional system,
which is finite in the direction perpendicular 
to the $x$-direction [the transverse direction $\mbox{\boldmath $\rho$}$].
The extension can be done following along almost 
the same lines as those for the one-dimensional system.
This is because the correspondence of the 
the imaginary part and the discontinuity   
of the Green's function 
holds also for the quasi-one-dimensional system, Eq.\ (\ref{eq:Im_GQ}), 
owing to the time reversal symmetry. 
So, the analysis of the Green's function used 
in Sec.\ \ref{sec:trans}--\ref{sec:Kubo} is also available here,
and thus we will present only the outline of the extension.
In Sec.\ \ref{subsec:trans}, 
we introduce the transmission and reflection coefficients 
for a number of scattering channels, Eq.\ (\ref{eq:transQ}),
making use of the the single-particle Green's function, 
and examine the unitarity condition for the coefficients 
generalizing the discussion in Sec.\ \ref{sec:trans}.
In Sec.\ \ref{subsec:cond}, 
we consider the contributions of the vertex correction 
to the dc conductance in the quasi-one-dimensional system
generalizing the calculations performed in Sec.\ \ref{sec:Kubo},
and as a result the conductance is expressed 
in terms of the single-particle Green's function or 
the transmission coefficient; 
Eqs.\ (\ref{eq:cond1Q}) or (\ref{eq:cond2Q}).

\subsection{Scattering Coefficients}
\label{subsec:trans}

In this subsection, 
we generalize the definition of 
the transmission and reflection coefficients discussed 
in Sec.\ \ref{sec:trans} to the quasi-one-dimensional system. 
We now introduce a set of normal modes $\chi_a(\mbox{\boldmath $\rho$})$ 
satisfying the Schr\"{o}dinger equation for 
the transverse direction in the leads  
\begin{equation}
\left[\, -  {1 \over 2m}\, 
{\partial^2 \over \partial \mbox{\boldmath $\rho$}^2} \ + \ 
V_{c}(\mbox{\boldmath $\rho$}) \,\right] \chi_a(\mbox{\boldmath $\rho$})
\ =\ \epsilon_a\, \chi_a(\mbox{\boldmath $\rho$}) \;,
\label{eq:mode}
\end{equation}
where $\, a$ is the subband index, $\, \epsilon_a$ 
is the subband energy, and $\,V_c(\mbox{\boldmath $\rho$})$ 
is confinement potential 
for the transverse direction which is perpendicular to the $x$-direction.
In what follows we choose the normal modes 
to be real, i.e., $\chi_a^{\phantom{*}}(\mbox{\boldmath $\rho$}) =
\chi_a^{*}(\mbox{\boldmath $\rho$})$.
This is possible because of 
the time reversal symmetry of Eq.\ (\ref{eq:mode}), and
is convenient for the discussion in the following.
The free Green's function is diagonal in the subband indices
\begin{equation}
G_0(\mbox{\boldmath $r$},\mbox{\boldmath $r$}'; z)
= \sum_a \chi_a(\mbox{\boldmath $\rho$})\,
\chi_a(\mbox{\boldmath $\rho$} ')
  \, G_a^{(0)}(x,x'; z) \;,
\end{equation}
where $\,\mbox{\boldmath $r$} = (x,\,\mbox{\boldmath $\rho$})$, 
$\,\mbox{\boldmath $r$} ' = (x',\,\mbox{\boldmath $\rho$} ')$, 
and $G_a^{(0)}(x,x'; z)$ is the free Green's function 
for the one dimension, which is obtained 
by replacing the chemical potential $\,\mu$ by $\,\mu-\epsilon_a$ 
in the right-hand side of Eqs.\ (\ref{eq:G0_ret}) 
or (\ref{eq:G0_Matsubara})-(\ref{eq:phi}). 
When the Coulomb interaction and 
the scattering potential are restricted 
for the electrons staying in the central region $\,0 \leq x \leq L\,$, 
the Dyson equation is written, by generalizing Eq.\ (\ref{eq:Dyson_T}),
as 
\begin{eqnarray}
  G_{ba}(x,x'; z) &=& G_b^{(0)}(x,x'; z) \, \delta_{ba}
 \nonumber \\
& & + \int_{0}^{L} \!\!\! \int_{0}^{L}  \!\! dx_1 dx_2  
 \, G_b^{(0)}(x,x_1; z)\ 
 {\cal T}_{ba}(x_1,x_2; z)\ G_a^{(0)}(x_2,x'; z) \;.
  \label{eq:Dyson_TQ}
\end{eqnarray}
Here the subscript $ba$ denotes 
the subband indices defined by
\begin{equation}
G_{ba}(x,x';z) \equiv
    \displaystyle \int \!\!\! \displaystyle\int \!\! 
   d\mbox{\boldmath $\rho$} d\mbox{\boldmath $\rho$}' \, 
\chi_b(\mbox{\boldmath $\rho$} )\, 
  G(\mbox{\boldmath $r$},\mbox{\boldmath $r$}'; z)\,
\chi_a(\mbox{\boldmath $\rho$}') \;.
\label{eq:subindex}
\end{equation}
Because of the time-reversal symmetry,  
the relation corresponding to Eq.\ (\ref{eq:symmetry}) holds 
also for the quasi-one-dimensional system as 
$G(\mbox{\boldmath $r$}',\mbox{\boldmath $r$}; z)
= G(\mbox{\boldmath $r$},\mbox{\boldmath $r$}'; z)$. 
Thus using Eq.\ (\ref{eq:subindex}), we obtain 
\begin{equation}
G_{ab}(x,x'; z) = G_{ba}(x',x; z)\;. 
\label{eq:symQ}
\end{equation}
Therefore, as in the one-dimensional case,
the discontinuity of the Green's function 
at the real axis in the complex $z$-plane corresponds to 
the imaginary part;
\begin{equation}
 \left[\, G_{ab}(x,x'; \varepsilon+i0^+)
- G_{ab}(x,x'; \varepsilon-i0^+) \,\right] / 2i \ = \
\mbox{Im}\, G_{ab}(x,x'; \varepsilon+i0^+) \;.
\label{eq:Im_GQ}
\end{equation}
Owing to this property, 
we can obtain the imaginary part using 
the perturbation analysis for extracting 
the singular contributions.

When both $x$ and $x'$ belong the leads,
the asymptotic form of the full Green's function 
can be written explicitly making use of Eq.\ (\ref{eq:Dyson_TQ}).
So, the equation corresponding to Eq.\ (\ref{eq:G_AB}) 
is written by taking the values of $x'$ and $x$ to be 
$\,x'< 0\,$ and $\,L<x\,$; 
\begin{equation}
G_{ba}(x, x';i0^+) = 
 \, e^{i\,k_b x} 
       \left[\, \delta_{ba} \, - \, {i \over v_b}\, {\cal T}_{ba:k_b k_a}(i0^+)
       \,\right]
 \,  \left(-{i \over v_a}\right) \,e^{-i\,k_a x'}\ 
\;,
  \label{eq:G_ABQ}   
\end{equation}
where the frequency is chosen to be $\, z= i0^+$, 
$k_a \equiv \sqrt{2m(\mu-\epsilon_a)}$ is the wave number 
of single-particle-like excitation at Fermi energy, 
and $v_a \equiv k_a/m$ is the velocity.
In Eq.\ (\ref{eq:G_ABQ})   
the initial and final states, $a$ and $b$, 
are assumed to be propagating modes with real values of
the wave number.
${\cal T}_{ba:k_b k_a}(z)$ is the Fourier transform of
${\cal T}_{ba}(x_1,x_2; z)$ with respect to $x_1$ and $x_2$ 
[see Eq.\ (\ref{eq:Tkk'})]. 
Similarly, 
the equation corresponding to Eq.\ (\ref{eq:G_AA}) is written  
by taking the positions to be $\,x'< x<0\,$; 
\begin{equation}
G_{ba}(x, x';i0^+) = 
       \left[\, \delta_{ba} \, e^{i\,k_a x }\, - \, {i \over v_b}\, 
    {\cal T}_{ba:-k_b k_a}(i0^+) \ e^{-i\,k_b x }
       \,\right]
 \,  \left(-{ i \over v_a}\right)\, e^{-i\,k_a x'}  
\;.
  \label{eq:G_AAQ}   
\end{equation}
Since the retarded Green's function $G_{ba}(x, x';i0^+)$
represents a propagation of a single-particle-like excitation 
at the Fermi energy as discussed in Sec.\ \ref{sec:trans}, 
it seems natural to define the scattering coefficients 
making use of Eqs.\ (\ref{eq:G_ABQ}) and (\ref{eq:G_AAQ}) as 
\begin{eqnarray} 
  t_{ba} \ &\equiv& \  
        \left[\,\delta_{ba} 
          - {i \over v_b} \  {\cal T}_{ba:k_bk_a}(i0^+) \,\right]
         \sqrt{v_b \over v_a}    \;, 
\label{eq:transQ}
\\
  r_{ba} \ &\equiv&
      \ \  - {i \over v_b}   \ {\cal T}_{ba:-k_bk_a}(i0^+)\ 
         \sqrt{v_b \over v_a}
\;,
\label{eq:reflcQ}
\end{eqnarray} 
where a factor $\sqrt{v_b/v_a}$ is introduced for the normalization. 
Then the unitarity condition, 
$\sum_b' \left[\,|t_{ba}|^2 + |r_{ba}|^2 \,\right]=1$, 
is expressed in the form 
\begin{equation} 
  \mbox{Im}\, {\cal T}_{aa:k_ak_a}(i0^+)
   \ = \  -\, 
        {\sum_b}' {1\over 2 v_b}
          \left[\, |{\cal T}_{ba:k_bk_a}(i0^+)|^2 + 
             |{\cal T}_{ba:-k_bk_a}(i0^+)|^2 \,\right] \;,
 \label{eq:opticalQ}
\end{equation}
where the primed sum runs over the propagating modes.
At $T=0$, the unitarity condition holds for 
the excitation at the Fermi energy  
if the imaginary part of the proper self-energy is 
zero;
\begin{equation}
\mbox{Im}\, \Sigma_{ab}(x,x'; \,i0^+) =0 \;.
\label{eq:Im_SQ}
\end{equation}
As in the one-dimensional case, 
Eq.\ (\ref{eq:opticalQ}) can be derived by using the operator 
identity Eq.\ (\ref{eq:OperatorID}). 
In the present case the matrix element of the operator 
$\mbox{\boldmath ${\cal T}$}$ is defined by 
$\langle b, k_b |\mbox{\boldmath ${\cal T}$}|a,k_a \rangle 
\equiv {\cal T}_{ba:k_bk_a}(i0^+)$, 
and the product of the operators is understood to be an integral 
for the motion in the $x$ direction and 
a sum over the subband index. 
The operator for the proper self-energy 
$\mbox{\boldmath $\Sigma$}$ is Hermitian owing to Eq.\ (\ref{eq:Im_SQ}).
Thus Eq.\ (\ref{eq:opticalQ}) can be obtained 
from the diagonal element of the identity in the wave-number indices,    
which is written by generalizing 
 Eq.\ (\ref{eq:OperatorID_diagonal}) as
\begin{equation}
{\cal T}_{aa:k_ak_a}^{\phantom{*}}(i0^+) 
 - {\cal T}_{aa:k_ak_a}^{*}(i0^+) 
\ = \  - \sum_b \int_{-\infty}^{+\infty} \! {dk'\over 2\pi} \ 
  {\cal T}_{ba:k'k_a}^{*}(i0^+)\  
 2 \pi i  \delta(k'^2/2m + \epsilon_b -\mu) \ {\cal T}_{ba:k'k_a}(i0^+) 
\;. 
\label{eq:OperatorID_diagonalQ} 
\end{equation} 
Because of the energy conservation due to the delta function, 
the sum over $\,b\,$ is restricted to the propagating modes 
and thus we obtain Eq.\ (\ref{eq:opticalQ}). 
We note that Eq.\ (\ref{eq:Im_SQ}) can also 
be derived in the perturbation theory 
by extending the calculation of 
the imaginary part performed in the Sec.\ \ref{sec:self-energy};  
by including sums over subband indices for intermediate states      
and using Eq.\ (\ref{eq:Im_GQ}).

\subsection{Conductance}
\label{subsec:cond}

Next we consider the expression of the conductance 
by extending the calculation performed in Sec.\ \ref{sec:Kubo}.
In the present case, the operator for the current in the $x$ direction 
is given by 
\begin{equation}
 J(x)\  = \   \sum_{\sigma} 
 \int\!\! d\mbox{\boldmath $\rho$}\   
{e \over 2 m i} 
 \left[\,
 \psi^{\dagger}_{\sigma}(x, \mbox{\boldmath $\rho$})\ 
{\partial \psi_{\sigma}(x, \mbox{\boldmath $\rho$}) \over \partial x} 
\ - \
{\partial \psi^{\dagger}_{\sigma}(x, \mbox{\boldmath $\rho$}) 
\over \partial x}\ 
\psi_{\sigma}(x, \mbox{\boldmath $\rho$})
\,\right]  \;. 
\end{equation} 
When the electric field is applied to the central region 
along the $x$ direction, 
the dc conductance within the Kubo formalism is given by   
\begin{eqnarray} 
& &
g \ = \ \lim_{\omega \to 0}\, 
 { \widetilde{K}(x, x';\omega+i0^+) 
- \widetilde{K}(x, x';i0^+)  \over  i \omega} \;, \\
\label{eq:KuboQ} 
& &  \widetilde{K}(x, x';\omega+i0^+) \equiv
i \int_0^{\infty} \! \! dt \,
\langle\,  \left[\, J(x, t)\,, J(x',0) \,\right]\, \rangle 
\ e^{i\,(\omega +i0^+)\, t } 
\;. \label{eq:K_RQ}  
\end{eqnarray}
where $x$ and $x'$ are arbitrary 
owing to the current conservation.
In the Kubo formalism 
the dc conductance is obtained from  
the $\omega$-linear imaginary part of $\,\widetilde{K}(x,x';\omega+ i0^+ )$.
In addition, the $\omega$-linear imaginary part 
corresponds to the $\,\nu\ \mbox{sgn}\, \nu\,$ term  
in the Matsubara function $\,\widetilde{K}(x,x'; i\nu )\,$
because the relation Eq.\ (\ref{eq:Im_K}) 
also holds for $\,\widetilde{K}$ 
owing to the symmetry property 
$\,\widetilde{K}(x,x';z)=\widetilde{K}(x',x;z)$.
Therefore, the conductance is obtained by 
extracting the $\,\nu\,\mbox{sgn}\,\nu\,$ term using 
the diagrammatic analysis. 
At $\,T=0$, 
the equations corresponding to Eqs.\ (\ref{eq:K_nu})--(\ref{eq:K_nu_b})
are written as
\begin{eqnarray}
\widetilde{K}(x,x';i \nu) \  
&=& \ \widetilde{K}^{(a)}(x,x';i \nu) \ 
+ \ \widetilde{K}^{(b)}(x,x';i \nu) \quad , 
\label{eq:K_nuQ} \\
\widetilde{K}^{(a)}(x,x';i \nu) &=& 
- \left({e \over 2 m i}\right)^2 
\left\{ {\partial / \partial x_4}\,
 - \,{\partial / \partial x_1} \right\} 
\left\{ {\partial / \partial x_3}\, 
 - \,{\partial / \partial x_2} \right\}
\nonumber  \\
& &
\times  
\sum_{\sigma} \sum_{ab}
\int_{-\infty}^{+\infty} \! {d\varepsilon \over 2 \pi} \ 
\biggl.
G_{ab}(x_3,x_1;i\varepsilon) \,  G_{ba}(x_4,x_2;i\varepsilon + i\nu)\,
\biggr|_{\scriptstyle x_1, x_4 \to x \atop \scriptstyle \ x_2, x_3 \to x'} 
\;\;,  
\label{eq:K_nu_aQ} \\ 
\widetilde{K}^{(b)}(x,x';i \nu) &=& 
- \left({e \over 2 m i}\right)^2 
\left\{ {\partial/\partial x_4}\,
 - \,{\partial/\partial x_1} \right\} 
\left\{ {\partial/\partial x_3}\, 
 - \,{\partial/\partial x_2} \right\}
\nonumber  \\
& & \times \sum_{\sigma \sigma '} 
         \sum_{ab} \sum_{\scriptstyle a_1 a_2 \atop \scriptstyle a_3 a_4} 
\int_{-\infty}^{+\infty} \! {d\varepsilon d\varepsilon' \over (2 \pi)^2}  
 \int_{0}^{L} \!\!\!\!\! \cdots \!\! \int_{0}^{L} 
   \prod_{i=1}^4 dy_i \, 
 G_{a_1 b}(y_1,x_1;i\varepsilon) \,  G_{b a_4}(x_4,y_4;i\varepsilon + i\nu)\,
\nonumber  \\
& & \times 
\Gamma_{\sigma \sigma '; \sigma' \sigma}^{a_4 a_3; a_2 a_1}
(y_4, y_3; y_2, y_1: 
i \varepsilon + i \nu, i \varepsilon';
i \varepsilon' + i \nu, i \varepsilon) 
\nonumber  \\
& & \times
\biggl.
G_{a a_3}(x_3,y_3;i\varepsilon') \,  G_{a_2 a}(y_2,x_2;i\varepsilon' + i\nu)\,
\biggr|_{\scriptstyle x_1, x_4 \to x \atop \scriptstyle \ x_2, x_3 \to x'} 
\;\;.  
\label{eq:K_nu_bQ}
\end{eqnarray}
Here superscripts of the vertex function denote the subband indices defined by
\begin{equation}
\Gamma_{\sigma \sigma '; \sigma' \sigma}^{a_4 a_3; a_2 a_1}
(x_4, x_3; x_2, x_1) 
\equiv  \int \! 
   \prod_{i=1}^4 d\mbox{\boldmath $\rho_i$} \ 
\chi_{a_4}(\mbox{\boldmath $\rho_4$} )\, 
\chi_{a_3}(\mbox{\boldmath $\rho_3$} )\, 
\Gamma_{\sigma \sigma '; \sigma' \sigma}
(\mbox{\boldmath $r_4$}, \mbox{\boldmath $r_3$}; 
\mbox{\boldmath $r_2$}, \mbox{\boldmath $r_1$}) 
\chi_{a_2}(\mbox{\boldmath $\rho_2$})\,
\chi_{a_1}(\mbox{\boldmath $\rho_1$}) \;.
\end{equation}
In what follows we 
choose $\,x$ to be in the right lead and $\, x'$ to be 
in the left lead, and thus $x'<0<y_i<L<x$ with $i=1,\ldots,4$.
Then, the derivative with respect to $x_1\ldots,x_4$ 
can be performed explicitly making use of the asymptotic form 
of the Green's function, 
and the equations corresponding 
to Eqs.\ (\ref{eq:bubble}) and (\ref{eq:vertex}) can be obtained as
\begin{eqnarray}
\widetilde{K}^{(a)}(x,x';i \nu) &=& 
- \left({e \over 2 m }\right)^2 
\sum_{\sigma} \sum_{ab}
\int_{-\infty}^{+\infty} \! {d\varepsilon \over 2 \pi} \ 
\{ \kappa_b(i\varepsilon + i\nu) - \kappa_b(i\varepsilon) \}\,
\{ \kappa_a(i\varepsilon + i\nu) - \kappa_a(i\varepsilon) \}
\nonumber \\
& & \times
G_{ab}(x',x;i\varepsilon) \,  G_{ba}(x,x';i\varepsilon + i\nu)\,
\;\;,  
  \label{eq:bubbleQ} \\
\widetilde{K}^{(b)}(x,x';i \nu) &=& 
- \left({e \over 2 m }\right)^2 
\sum_{\sigma \sigma '} 
         \sum_{ab} \sum_{\scriptstyle a_1 a_2 \atop \scriptstyle a_3 a_4} 
\int_{-\infty}^{+\infty} \! {d\varepsilon d\varepsilon' \over (2 \pi)^2}  
 \int_{0}^{L} \!\!\!\!\! \cdots \!\! \int_{0}^{L} 
   \prod_{i=1}^4 dy_i \, 
\nonumber\\
& &
\times
\{ \kappa_b(i\varepsilon + i\nu) - \kappa_b(i\varepsilon) \}\,
 G_{a_1 b}(y_1,x;i\varepsilon) \,  G_{b a_4}(x,y_4;i\varepsilon + i\nu)\,
\nonumber  \\
& & \times 
\Gamma_{\sigma \sigma '; \sigma' \sigma}^{a_4 a_3; a_2 a_1}
(y_4, y_3; y_2, y_1: 
i \varepsilon + i \nu, i \varepsilon'; 
i \varepsilon' + i \nu, i \varepsilon) 
\nonumber  \\
& & \times
\{ \kappa_a(i\varepsilon' + i\nu) - \kappa_a(i\varepsilon') \}\,
G_{a a_3}(x',y_3;i\varepsilon') \,  G_{a_2 a}(y_2,x';i\varepsilon' + i\nu)
\;.
\label{eq:vertexQ}
\end{eqnarray}
Here $\kappa_a(i\varepsilon)$ denotes the quantity 
which is given by replacing $\,\mu$ by $\,\mu-\epsilon_a$ 
in Eqs.\ (\ref{eq:kappa}) and (\ref{eq:phi}). 
At $T=0$, the contributions of the vertex correction, Eq.\ (\ref{eq:vertexQ}), 
to the dc conductance becomes zero as that in the one-dimensional case. 
This is because the {\em current vertex\/} 
$\{ \kappa_a(i\varepsilon' + i\nu) - \kappa_a(i\varepsilon') \}$
is zero for $\nu =0$. 
This feature can also be understood in terms of 
the three point vertex function for $J(x')$, 
which represents the renormalization of the current 
or the effects of the back flow as discussed 
in the last part of Sec.\ \ref{sec:Kubo} 
[see Fig.\ \ref{fig:currentV}]. 
Consequently, the dc conductance is determined by 
the $\nu\,\mbox{sgn}\, \nu$ term 
in the right-hand side of Eq.\ (\ref{eq:bubbleQ}), and   
the expression of the dc conductance for $T=0$ is obtained 
by taking the asymptotic limit 
$x' \to -\infty $ and $x \to \infty$  
in order to suppress the contribution of the evanescent modes 
with imaginary values of the wave number,\cite{Fisher} 
and reinserting $\,\hbar$;
\begin{equation}
 g  =  {e^2 \over \pi\hbar} {\sum_{ba}} '  
                     v_a v_b  \left|\, G_{ba}(x,x';i0^+)\, \right|^2  
 \label{eq:cond1Q}   \;,
\end{equation}
where the primed sum runs over the propagating modes. 
This expression can be written in terms of 
the transmission coefficient $t_{ab}$ defined by Eqs.\ (\ref{eq:G_ABQ}) 
and (\ref{eq:transQ}) as 
\begin{equation}
 g =  {2  e^2 \over h}  {\sum_{ba}}' \left|t_{ba} \right|^2  \;.
\label{eq:cond2Q}
\end{equation}
These are the results obtained with the perturbation theory.

\section{SUMMARY}
\label{sec:summary}
We have studied the conductance 
for the current through a finite central region 
which is connected to semi-infinite noninteracting leads 
on the left and the right. 
Assuming the Coulomb interaction $\,U$ and 
the normal scattering potential with time reversal symmetry $\,V$
can be switched on only for the electrons in the central region, 
we have stressed the study of the relation 
between the conductance in the Kubo formalism 
and the transmission coefficient 
introduced by making use of the single-particle Green's function.
In order to calculate the contributions 
of the vertex correction for the dc conductance, 
we have used the diagrammatic analysis 
for the perturbation expansion with respect to the Coulomb interaction,  
which was originally 
applied to the impurity Anderson model for the Kondo alloy. 
At $\,T=0$, 
the contributions of the vertex correction 
to the dc conductance become zero 
if the currents are measured in the leads.
This is caused by the fact that the asymptotic form of the Green's function
in the leads is given by a simple superposition of the incident,
transmitted, and reflected waves.
Consequently, the conductance is expressed 
in a Landauer-type form using the transmission coefficient 
for single-particle-like excitation at the Fermi energy 
[see Eq.\ (\ref{eq:cond2})]. 
At $T=0$, the scattering coefficients are confirmed to satisfy 
the unitarity condition owing to a Fermi-liquid property  
$\, \mbox{Im}\,\Sigma(x,x';i0^+) = 0$,
and this property can be derived with the perturbation theory. 
The results are generalized to a quasi-one-dimensional system 
with a number of scattering channels [see Eq.\ (\ref{eq:cond2Q})]. 
The analysis used in the present study can also be applied to 
a tight-binding model, 
and the similar results can be derived for 
the model on a lattice.\cite{ao5}

Throughout the present study,
we have assumed that the perturbation expansion 
in the Coulomb interaction is valid.
For instance, in the limit the size of the central region $L \to 0$, 
the model is reduced to an Anderson-Wolff impurity 
and the perturbation theory is valid for all values of $\,U$.
Although it is not evident that the perturbation theory
is valid for all values of $\,U$, $\,V$, and $\,L$ in the general case,
the hypothesis may be valid at least 
for a {\em finite region\/} of the parameter space of the Hamiltonian 
where $\,U$, $\,V$, and $\,L$ are small enough.

\acknowledgments

We would like to thank H. Ishii for valuable discussions.
Allocation of computer time 
at the Supercomputer Center, 
Institute for Solid State Physics, University Tokyo, 
and at the Computer Center of the Institute for Molecular Science, 
Okazaki National Research Institutes  
are gratefully acknowledged.
This work was partly supported by a Grant-in-Aid 
from the Ministry of Education, Science, and Culture.

\end{document}